\documentclass[preprint,12pt]{elsarticle}

\usepackage{amsmath,amsfonts,amsthm,amssymb,paralist,subfigure,graphicx,amsbsy,float,epsfig,color}
\usepackage{cuted,mathtools,lipsum}
\usepackage{bm}
\usepackage{algorithm2e,algorithmic}
\usepackage{pifont}
\usepackage{hyperref}       
\hypersetup{
	colorlinks=true,
	linkcolor=cyan,
	filecolor=mnodea,
	urlcolor=cyan,
	citecolor=lime,
}
\usepackage{url}            
\usepackage{booktabs}       
\usepackage{amsfonts,amsmath,amssymb}       
\usepackage{amsthm}     
\usepackage{nicefrac}       
\usepackage{microtype}      
\usepackage{lipsum}
\usepackage{graphicx}
\usepackage{dsfont}
\usepackage{enumitem}
\usepackage{stmaryrd,url}

\usepackage{amsthm}
\newtheorem{lem}{Lemma}
\newtheorem{ass}{Assumption}
\newtheorem{thm}{Theorem}

\newtheorem{rem}{Remark}

\def\mb{\mathbf}

\def\mc{\mathcal}

\journal{Systems and Control Letters}

\begin{document}

\begin{frontmatter}

\title{Distributed Target Tracking  based on Localization with Linear Time-Difference-of-Arrival Measurements:\\ A Delay-Tolerant Networked Estimation Approach}

\author[Sem]{Mohammadreza Doostmohammadian}
\affiliation[Sem]{Faculty of Mechanical Engineering, Semnan University, Semnan, Iran, doost@semnan.ac.ir.}

\author[Aalto,TC]{Themistoklis Charalambous}
\affiliation[Aalto]{School of Electrical Engineering, Aalto University, Espoo, Finland, name.surname@aalto.fi.}
\affiliation[TC]{School of Electrical Engineering, University of Cyprus, Nicosia, Cyprus, surname.name@ucy.ac.cy.}


\begin{abstract}
	This paper considers target tracking based on a beacon signal's time-difference-of-arrival (TDOA) to a group of cooperating sensors. The sensors receive a reflected signal from the target where the time-of-arrival (TOA) renders the distance information. The existing approaches include: (i) classic centralized solutions which gather and process the target data at a central unit, (ii) distributed solutions which assume that the target data is observable in the dense neighborhood of each sensor (to be filtered locally), and (iii) double time-scale distributed methods with high rates of communication/consensus over the network. This work, in order to reduce the network connectivity in (i)-(ii) and communication rate in (iii), proposes a distributed single time-scale technique, which can also handle heterogeneous constant data-exchange delays over the static sensor network. This work assumes only \textit{distributed observability} (in contrast to local observability in some existing works categorized in (ii)), i.e., the target is observable globally over a (strongly) connected network. The (strong) connectivity further allows for \textit{survivable network} and \textit{$q$-redundant observer design}. Each sensor locally shares information and processes the received data in its immediate neighborhood via local linear-matrix-inequalities (LMI) feedback gains to ensure tracking error stability. The same gain matrix works in the presence of heterogeneous delays with no need of redesigning algorithms. Since most existing distributed estimation scenarios are linear (based on consensus), many works use \textit{linearization} of the existing \textit{nonlinear TDOA measurement models} where the output matrix is a function of the target position. As the exact target position is unknown, the existing works use \textit{estimated} position in the output matrix (and for the gain design) at every time step. This makes their algorithm more complex and less accurate. Instead, this work provides a \textit{modified} linear TDOA measurement model with a \textit{constant} output matrix that is independent of target position and more practical in distributed linear setups.
\end{abstract}

\begin{graphicalabstract}
	\includegraphics{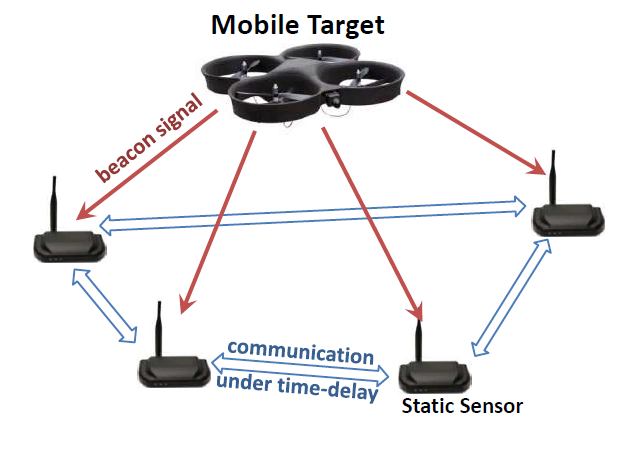}
\end{graphicalabstract}

\begin{highlights}
	\item Distributed single time-scale estimation strategies with no inner consensus loop for target tracking over sensor networks
	\item Delay-tolerant distributed tracking under heterogeneous  delays over the sensor network
	\item Computationally efficient linear TDOA-based measurements applicable for general distributed estimation strategies
	\item Relaxing local observability and sensor network connectivity for distributed estimation
\end{highlights}


\end{frontmatter}

\section{Introduction} \label{sec_intro}
Target tracking and localization  \cite{bar2004estimation,usman_loctsp:08,gustafsson2002particle,mohammadi2014distributed,ennasr2016distributed,ennasr2020time} finds potential applications in environmental surveillance, intelligent transportation systems, space explorations, and military applications, where the key problem is to localize and track a target via certain measurements.  Typical approaches for tracking include estimated time-of-arrival (TOA), direction-of-arrival (DOA),  received-signal-strength (RSS),  time-difference-of-arrival (TDOA), or a combination of these. We refer interested readers to \cite{dardari2015indoor} for some examples and surveys on these. In (centralized) TDOA-based tracking, the target broadcasts a beacon signal, and the TOAs of this beacon signal at all sensors acting as the anchors are given to a central unit to filter the data and localize the target. The TDOA values are converted to distance-based or range-difference measurements (based on the signal's propagation speed). TDOA-based distance measurements are adequately accurate for far-distance targets and are commonly used for outdoor as well as indoor tracking
\cite{ennasr2016distributed,ennasr2020time}.
In \textit{centralized} tracking, the TDOA-based measurement data taken by sensors are sent to the central unit (the reference node) which localizes the target by filtering the data \cite{simon_nonlin_filter,hartikainen2008optimal}.
Another interesting work \cite{gong2018auv} proposes a localization method via a mobile anchor node to replace a group of fixed anchor nodes for scalability purposes.

In contrast, recently \textit{distributed}  techniques via a \textit{network} of sensors are considered \cite{mohammadi2014distributed,ennasr2016distributed}. A group of \textit{smart sensors} (or \textit{agents}), embedded with data processing and communication devices, communicate and share data (e.g., via a wireless communication network), process these data locally, and make a decision on the target's location. Many such distributed scenarios are motivated by the emergence of parallel processing and edge computing \cite{zhao2020iot}, since the centralized case is prone to single-node-of-failure and high communication (or computation) burden at the central processing unit. Further, distributed setups are scalable, enabling the distribution of the processing/communication load across multiple nodes. In some distributed tracking literature,
each sensor shares sufficient information on the (local) observability of the target to its neighbors. This is widely adopted in networked estimation/filtering literature \cite{mohammadi2014distributed,ennasr2016distributed,ennasr2020time,jstsp14,rekabi2020distributed,Li2015Moura,kar2012distributed,das2015distributed,wu2019efficient,rastgar2018consensus,olfati2011collaborative,he2020secure,battistelli2014consensus,battilotti2021stability} based on consensus fusion algorithms.
Recall that nonlinear filtering methods \cite{gustafsson2002particle,simon_nonlin_filter,hartikainen2008optimal} are not well-developed in \textit{distributed} setups, where relevant concepts in terms of \textit{distributed observability} need to be further addressed. The existing distributed estimation methods are designed for linear systems/measurements, and thus, the ones related to the target tracking purposes use \textit{linearization} of the existing \textit{nonlinear measurements} instead \cite{ennasr2016distributed,ennasr2020time}, and this may degrade their accuracy and performance.

\begin{figure} [t]
\centering
\includegraphics[width=1.7in]{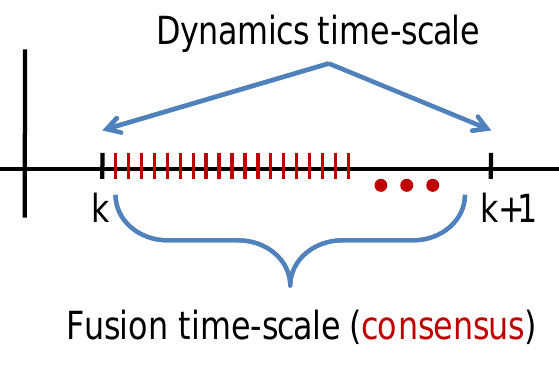}
\includegraphics[width=1.7in]{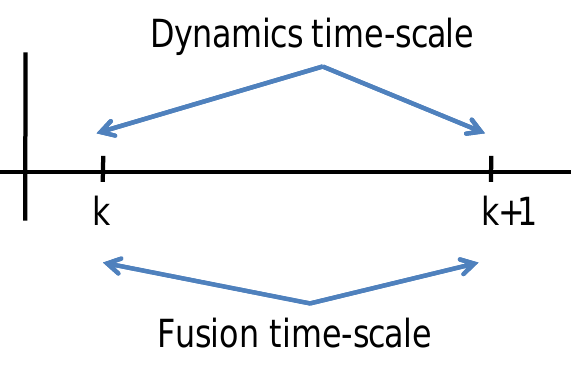}
\caption{Two types of consensus-based networked estimation mechanisms: (Left) double time-scale (DTS), and (Right) single time-scale (STS).
	The DTS scenario imposes many (fast) iterations of consensus and communications (in general,  more than the network diameter $d_n$) to perform estimation and address observability. In contrast, in the STS method, only one round of inter-sensor message-exchange is done per observation/system epoch.  }
\label{fig_fusion}
\end{figure}

Two types of consensus-updates are adopted in networked/distributed estimation: (i) double time-scale (DTS) method \cite{olfati2011collaborative, battistelli2014consensus, he2020secure,battilotti2021stability}, and (ii) single time-scale (STS) method \cite{mohammadi2014distributed,jstsp14,rekabi2020distributed,Li2015Moura,kar2012distributed,das2015distributed,wu2019efficient,rastgar2018consensus} (see Fig.~\ref{fig_fusion}).
In the STS approach, the time-scale of the target dynamics and estimation are the same, i.e., there is one epoch of averaging/consensus (and communication) among the sensors in between two consecutive steps ($k$ and $k+1$) of the system dynamics. In contrast, DTS filtering performs many epochs of averaging/consensus between $k$ and $k+1$. Therefore, the communication among the sensors needs to be implemented at a much faster rate than the sampling rate of the system dynamics. In this sense, the network connectivity, which is a real concern in the observability analysis of the STS strategy \cite{jstsp14}, becomes less relevant in the DTS method. This is because doing many iterations of information fusion over a sparse network in the DTS case is equivalent to the information exchange over an all-to-all network in the STS method \cite{usman_acc:11}.
Also, many works in the STS literature assume \textit{local} observability in the neighborhood of each sensor \cite{mohammadi2014distributed,rekabi2020distributed,Li2015Moura,kar2012distributed,das2015distributed,wu2019efficient}.
In many real-time applications with power constraints (e.g., mobile sensors with limited battery capacity) and low communication/computation budget (e.g., over a long-range spatially distributed network), the DTS technique is of less interest due to the high data-processing load and communication rate needed.
The STS methods, on the other hand, further require the calculation of certain distributed (or local) feedback gains \cite{khan2011coordinated,jstsp14,usman_cdc:10,han2018simple}, to enable the closed-loop estimator (observer) reach a desirable convergence rate. The local observability assumption, in many existing works, is ensured by dense connectivity over the network.

Other than consensus-based, some works consider diffusion-based strategies for distributed filtering and decision-making \cite{cattivelli2008diffusion}.
However, in contrast to \textit{dynamic} system tracking, these works track a \textit{static}  variable (also referred to as static linear state-space model), for example, to estimate the location of a source in a noisy environment \cite{tu2013distributed} or to solve similar regularized least-square problems \cite{cattivelli2008diffusion}.
Simultaneous estimation and anomaly detection (e.g., sensor faults or attacks) are also of interest.
There are two primary approaches for detection: (i) using the knowledge of system dynamics, referred to as linear dynamic state-space (LDS), and (ii) with no use of system dynamics, referred to as the static linear state-space (SLS) models. In general, existing physical-model-based (and observer-based) detection scenarios lie in the first category; see \cite{giraldo2018survey,dibaji2019systems} for surveys of existing detection techniques, which can be added to the distributed filtering (and tracking) to improve its resiliency in faulty environments. Similar to \cite{he2020secure}, the results of this paper allow for \textit{joint} distributed fault detection and isolation (FDI) techniques (following  \cite{tcns2021,icas21_attack}), and even designing redundant distributed observers as preventive techniques over faulty sensor networks. The need for linear TDOA measurements fed to distributed STS estimators with sparse connectivity further motivates this work.

\textbf{Main contributions}:
This paper adopts an STS networked estimation protocol
with no need for observability of the system (target) in the neighborhood of any sensor, i.e., with no assumption on local target observability at any sensor. The network connectivity required is such that every sensor \textit{gains (distributed) observability over the network}.
We prove error dynamics stability, even in the presence of heterogeneous fixed time-delays at different links, over strongly connected (SC) networks. Further, the SC network connectivity in this work is more relaxed, e.g., compared to \cite{ennasr2016distributed,usman_loctsp:08}. This enables $q$-redundant distributed observer design in this work which cannot be necessarily addressed in formation tracking of mobile sensors \cite{ennasr2020time} as their \textit{formation coordination} (to keep their distance constant) might be interrupted in case of link/node failure. In other words, our solution works for SC-designed networks after (up to) $q$ node or link removal (known as \textit{survivable network design} \cite{giraldo2018survey}) while this may cause unstable formation setup in \cite{ennasr2020time}. One can add \textit{localized} FDI at sensors to detect possible faults at the TDOA measurements, e.g., via both \textit{stateless} \cite{tcns2021} and \textit{stateful} \cite{icas21_attack} distributed methods based on probabilistic threshold design.
Recall that \textit{both estimation and detection scenarios are local with no need of any centralized coordinator (or decision-maker).} The proposed protocol, unlike \cite{ennasr2016distributed,mohammadi2014distributed,rekabi2020distributed,Li2015Moura,kar2012distributed,das2015distributed,wu2019efficient}, works under sparse network connectivity and possible bounded time-delay over the network.

We provide a \textit{linear} TDOA-based measurement model with a constant output matrix (independent of the target state) for distributed linear estimation which improves the linearized model in \cite{ennasr2020time}. This measurement model is used for general linear distributed estimation methods (both STS and DTS). We particularly compare our results with DTS scenarios \cite{olfati2011collaborative,battistelli2014consensus,he2020secure,battilotti2021stability} in terms of communication and processing rates. Recall that, from Fig.~\ref{fig_fusion}, the DTS methods perform many rounds of consensus and message-sharing epochs between two samples of system dynamics (to make the target observable at every time-step), and thus, require much faster communication and more processing load whereas this work requires only one epoch of consensus-update and communication between two (consecutive) steps of the system dynamics. This also works toward the delay-tolerance of our solution.  Over static sensor networks, this work addresses possible \textit{latency} over their data-exchange links. In case of mobile sensors, \textit{formation} scenarios are further needed to make the relative distance of the sensors fixed and constant \cite{fathian2019robust,ennasr2020time}, e.g., to reach final formation in fixed-time \cite{taes}; however, \textit{in the presence of time-delays}, such setup may not necessarily result in \textit{stable formation}.

To summarize, our approach  (i) requires less communication \textit{links} (compared to the other STS methods  \cite{mohammadi2014distributed,rekabi2020distributed,Li2015Moura,kar2012distributed,das2015distributed,wu2019efficient}), (ii) requires less communication/processing \textit{rate} (compared to the DTS filtering methods \cite{olfati2011collaborative, battistelli2014consensus,he2020secure,battilotti2021stability}), (iii) handles heterogeneous communication time-delays over the sensor network in contrast to \cite{ennasr2020time,ennasr2016distributed,mohammadi2014distributed,rekabi2020distributed,Li2015Moura,kar2012distributed,das2015distributed,wu2019efficient}, and (iv) addresses \textit{linear} TDOA-based measurements which is not considered in the existing distributed tracking literature \cite{ennasr2020time,ennasr2016distributed,mohammadi2014distributed}. The possibility of adding \textit{local} fault detection, and designing $q$-redundant observers further distinguishes our work. Note that our dynamic estimation differs from static models in \cite{tu2013distributed,cattivelli2008diffusion,cattivelli2008diffusion1} which disregard the system dynamics irrespective of observability concerns due to large number of observations more than the system size.

\textbf{Paper organization}:  Section~\ref{sec_frame} defines the target model and formulates the problem. Section~\ref{sec_measure} provides the measurement model for the TDOA-based tracking.  Section~\ref{sec_dist} proposes our (delay-tolerant) networked estimation and the error-stability analysis.  Section~\ref{sec_auxiliary} presents some discussions on $q$-redundant observability and fault detection. Section~\ref{sec_sim} provides simulations. Finally, Section~\ref{sec_con} concludes the paper. Proofs of lemmas and theorems are in the Appendix.

\section{The Framework}\label{sec_frame}
\subsection{Problem statement} \label{sec_prob}
The problem is to adopt a methodology to localize the target position (using the TDOA measurements). The classic approach is to send the TDOA measurement $\mb{y}_i$ of sensor $i$ to a central entity and use centralized estimation/filtering; this center tracks the target state $\mb{x}$. In this paper, each sensor tracks the target state in a distributed way. The solution to this problem mandates specific sensor network connectivity, better measurement models, and an appropriate local estimation/filtering design. The key point is to rely only on the information from the neighboring sensors $j \in \mc{N}_i$ along with the information of sensor $i$. In this direction, we define a network $\mc{G}=\{\mc{V},\mc{E}\}$ with the node set $\mc{V}=\{1,\dots,n\}$ representing the sensors and the link set $\mc{E}=\{(i,j)|j \in \mc{N}_i, i,j \in \mc{V}\}$ representing the communication channels among the sensors. The network is said to be (strongly) connected if for every pair of nodes $i,j$ there exists a path (sequence of links) from node $i$ to node $j$ and vice versa. 
\begin{rem}
		It should be clarified that this problem is not a leader-follower scenario. In a leader-follower scenario, a group of (typically mobile) follower agents track the state of a leader agent based on state information received from the leader. This paper instead considers a TDOA-based target tracking scenario, where the agents send/receive a beacon signal to/from the target and by sharing this TOA measurements along with their estimates with the neighboring agents localize the position of the target. Specific network design, localization procedure, and estimation techniques are needed to estimate the position of the target in a distributed way.   
\end{rem}
\begin{rem}
The existing distributed setups require the system-measurement pairs (i.e., the target dynamics and shared TOAs) to be observable at every sensor, either by adding more network connectivity or increasing the communication/consensus rate.
The idea in this paper is to develop a distributed tracking method with \textit{no need for local observability} by proper data-sharing and consensus-fusion among the sensors over a distributed estimation network. 
\end{rem}
In our setup, each sensor $i$ estimates the target state, denoted by $\widehat{\mb{x}}_i$, where only \textit{distributed observability} is needed. The objective is, then, two-fold: (i) to provide a \textit{linear time-invariant (LTI) measurement model} applicable in general existing distributed estimation setups, and  (ii) to develop a stable STS distributed estimator and proper design of the sensor network to ensure observability in the distributed sense. Our distributed setup further can tolerate data-sharing delays over the network and makes it possible to add simultaneous FDI and fault-tolerant network design.

\subsection{The target model}
In target tracking, the target's path is generally assumed unknown. In this sense, two main models are considered: Nearly-Constant-Velocity (NCV) and Nearly-Constant-Acceleration (NCA)\footnote{There are other dynamic models for target tracking as described in \cite{bar2004estimation},
for example, the Singer model; this model reduces to the NCV as the maneuvering time constant decreases, and it reduces to the NCA model as the maneuvering time constant increases \cite{bar2004estimation}. It should be noted that the methodology in this paper works for any target dynamics by adjusting the measurement parameters. Other than the NCV and NCA model, for example, the Singer model or other dynamic models proposed in \cite{bar2004estimation} can be adopted as the target dynamics. } \cite{bar2004estimation}. These models are commonly used in the literature \cite{roy2006target,gustafsson2002particle,ennasr2020time,ennasr2016distributed,bar2004estimation}. The state vector evolves as,

\begin{equation}  \label{eq_targ}
\mb{x}(k+1) = F\mb{x}(k)+G\mb{w}(k)
\end{equation}
where $\mb{x}(k)$ is the state-vector, $F$ and $G$, respectively, are the transition matrix and the input matrix, and $\mb{w}(k)$ is the process noise (as the random input), all at time-step $k$.

\textit{NCV model:}
In this model, the states represent the position and velocity of the target in 3D space. Define the target state vector as
$$\mb{x} = \left(p_x;p_y;p_z;\dot{p}_x;\dot{p}_y;\dot{p}_z\right),$$
where ``;'' is the column concatenation operator,
$p_x$, $p_y$, $p_z$ are positions and $\dot{p}_x$, $\dot{p}_y$, $\dot{p}_z$ are  velocities,  respectively, in $X,Y,Z$ coordinates. Then, the transition and input matrices are defined as \cite{bar2004estimation,gustafsson2002particle},
\begin{align}  \label{F_ncv}
F = \left(
\begin{array}{cc}
	\mb{I}_3 & T \mb{I}_3\\
	\mb{0}_3 & \mb{I}_3 \\ 	
\end{array} \right),~
G = \left(
\begin{array}{c}
	\frac{T^2}{2} \mb{I}_3 \\
	T\mb{I}_3 \\ 	
\end{array} \right)
\end{align} \normalsize
with $\mb{I}_3$ and $\mb{0}_3$ respectively representing the identity and zero matrix of size $3$, and $T$ as the sampling time interval
between two successive time-steps.

\textit{NCA model:}
In this model, the states are position, velocity, and acceleration in 3D space, and the state vector is defined as $$\mb{x} = \left(p_x;p_y;p_z;\dot{p}_x;\dot{p}_y;\dot{p}_z;\ddot{p}_x;\ddot{p}_y;\ddot{p}_z \right),$$
where $\ddot{p}_x$, $\ddot{p}_y$, $\ddot{p}_z$ are the accelerations, respectively, in $X,Y,Z$ directions. The transition and input matrix are respectively defined as \cite{bar2004estimation,gustafsson2002particle},
\begin{align}  \label{F_nca}
F = \left(
\begin{array}{ccc}
	\mb{I}_3 & T\mb{I}_3 & \frac{T^2}{2}\mb{I}_3\\
	\mb{0}_3 & \mb{I}_3 & T\mb{I}_3\\
	\mb{0}_3 & \mb{0}_3 & \mb{I}_3	
\end{array} \right), ~G = \left(
\begin{array}{c}
	\frac{T^2}{2} \mb{I}_3 \\
	T \mb{I}_3 \\
	\mb{I}_3	
\end{array} \right).
\end{align}
Both models are widely used in target-tracking scenarios.
\section{The Proposed Sensor Measurement Model} \label{sec_measure}
\begin{figure} [t]
\centering
\includegraphics[width=3in]{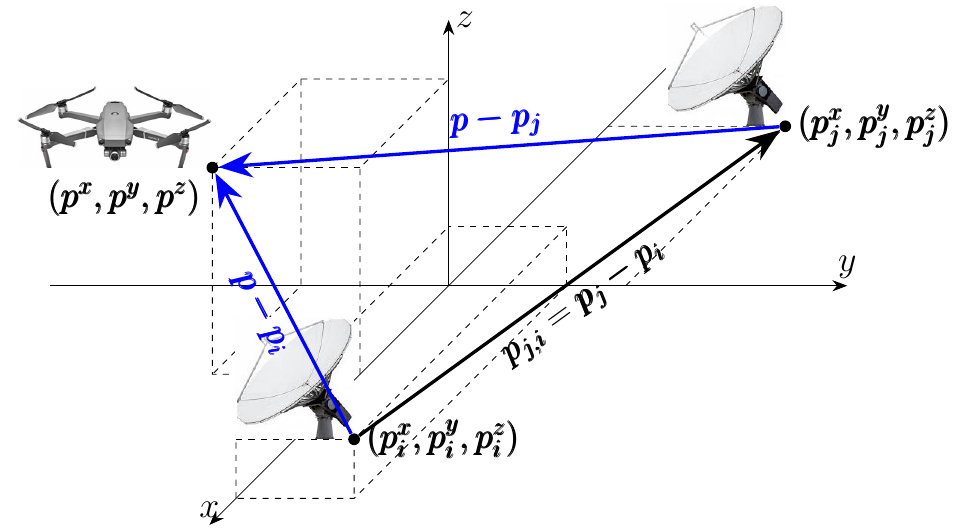}
\caption{This figure presents the target tracking setup in this paper: a group of geographically distributed static sensors (radars) receive a beacon signal from a target (drone) and locally (individually) track the target via distributed estimation and localized fault detection techniques. The measurements are based on the time-difference-of-arrival (TDOA) measurements described via Eq.~\eqref{eq_h_simp} and Eq.~\eqref{eq_tdoa}.  }
\label{fig_radar_drone}
\end{figure}

Assume a network of $n$ sensors with  $\mb{p}_i = (p_{x,i};p_{y,i};p_{z,i})$ as the position of sensor $i$ in the tracking field.
Every sensor receives a beacon signal \textit{with known propagation speed $c$} from the mobile target. At time-instant $k$, sensor $i$ saves the TOA of the received signal $t_i=\frac{1}{c}\|\mb{p}(k)-\mb{p}_i(k)\|$ and shares this along with its position with the set of its direct neighboring sensors $\mc{N}_i$ (this neighboring set $\mc{N}_i$ includes few other sensors and not necessarily all other sensors). Therefore, each sensor possesses a list of TOAs of the neighboring sensors at each time instant. Subtracting these TOAs from its own TOA measurement, the sensor finds the TDOA measurements.
The TDOAs are converted and interpreted as range-difference information.
The TDOA measurement of sensor $i$ at time $k$ is defined as,
\begin{align} \label{eq_y_nonlin}
\mb{y}_i(k) = h_i(k) +\bm \nu_i(k),
\end{align}
where $\mb{y}_i$ is the measurement vector, $\bm \nu_i$ is the zero-mean (white) measurement noise at sensor $i$, and $h_i$ is a function of the target position ${\mb{p}=(p_{x};p_{y};p_{z})}$ (and target state $\mb{x}$)
defined as,
\begin{align}
h_i(\mb{x}(k))=\left(
\begin{array}{c}
	h_{i,j_1}(\mb{x}(k))\\
	h_{i,j_2}(\mb{x}(k))  \\
	\vdots \\
	h_{i,j_{|\mc{N}_i|}}(\mb{x}(k))	
\end{array} \right),
\end{align}
with $\mc{N}_i = \{j_1,j_2,\dots,j_{|\mc{N}_i|}\}$ as the set of neighbors of $i$ and,
\begin{align} \label{eq_tdoa}
h_{i,j}(\mb{x}(k)) = \|\mb{p}(k)-\mb{p}_i(k)\|-\|\mb{p}(k)-\mb{p}_{j}(k)\|.
\end{align}
with $\|.\|$ as the Euclidean norm. The nonlinear measurement model \eqref{eq_y_nonlin} can be linearized as \cite{bar2004estimation},
\begin{align} \label{eq_y_lin}
\mb{y}_i(k) = H_i(k)\mb{x}(k) + \bm \nu_i(k).
\end{align}
Dropping variable $k$ (for notation simplicity) in the sequel, the time-dependent $H_i(k)$ for the NCV model is denoted by $\overline{H}_i$,
\begin{align} \label{eq_h_ncv}
\overline{H}_i=\left(
\begin{array}{cccccc}
	\frac{\partial h_{i,j_1}}{\partial p^{x}} & \frac{\partial h_{i,j_1}}{\partial p^{y}} & \frac{\partial h_{i,j_1}}{\partial p^{z}} &0& 0& 0\\
	\vdots & \vdots & \vdots & \vdots & \vdots & \vdots\\
	\frac{\partial h_{i,j_{|\mc{N}_i|}}}{\partial p^{x}} & \frac{\partial h_{i,j_{|\mc{N}_i|}}}{\partial p^{y}} & \frac{\partial h_{i,j_{|\mc{N}_i|}}}{\partial p^{z}} &0& 0& 0
\end{array} \right),
\end{align}
where,
\begin{align}
\frac{\partial h_{i,j}}{\partial p_{x}} = \frac{p_x-p_{x,i}}{\|\mb{p}-\mb{p}_i\|}-\frac{p_x-p_{x,j}}{\|\mb{p}-\mb{p}_{j}\|},\\
\frac{\partial h_{i,j}}{\partial p_{y}} = \frac{p_y-p_{y,i}}{\|\mb{p}-\mb{p}_i\|}-\frac{p_y-p_{y,j}}{\|\mb{p}-\mb{p}_{j}\|},\\ \label{eq_hij}
\frac{\partial h_{i,j}}{\partial p_{z}} = \frac{p_z-p_{z,i}}{\|\mb{p}-\mb{p}_i\|}-\frac{p_z-p_{z,j}}{\|\mb{p}-\mb{p}_{j}\|}.
\end{align}
Similar measurement models can be given for the NCA model. The measurement model is better illustrated in Fig.~\ref{fig_radar_drone}. Note that the traditional measurement model in \eqref{eq_h_ncv}-\eqref{eq_hij} includes \textit{state-dependent} terms, i.e., they are a function of $\mb{x}$ (or more accurately $\mb{p}$) as the changing state of the target.
In this work, we propose the following modified TDOA measurement model,

\small\begin{align} \nonumber
h_{i,j}(\mb{x}(k)) &= \frac{1}{2}\Big(\|\mb{p}(k)-\mb{p}_i(k)\|^2-\|\mb{p}(k)-\mb{p}_{j}(k)\|^2\Big) + \nu_{i,j}(k) \\ \nonumber
&= \frac{1}{2} \Big((p_{x,j}-p_{x,i})(2p_x-p_{x,j}-p_{x,i}) \\ \nonumber
& ~ +(p_{y,j}-p_{y,i})(2p_y-p_{y,j}-p_{y,i}) \\ \nonumber
& ~ + (p_{z,j}-p_{z,i})(2p_z-p_{z,j}-p_{z,i}) \Big) + \nu_{i,j}(k) \\
&= H_i \mb{x}(k) - \frac{1}{2}(\|\mb{p}_j\|^2 - \|\mb{p}_i\|^2) + \nu_{i,j}(k)
\label{eq_h_simp0}
\end{align}	\normalsize
where,
\begin{align}
\label{eq_h_simp}
H_i=\left(
\begin{array}{cccccc}
	p_{x,j,i} & p_{y,j,i} &  p_{z,j,i} & 0 & 0 & 0\\
	\vdots & \vdots & \vdots & \vdots & \vdots & \vdots\\
	p_{x,j_{|\mc{N}_i|},i} & p_{y,j_{|\mc{N}_i|},i} &  p_{z,j_{|\mc{N}_i|},i} & 0 & 0 & 0\\
\end{array} \right),
\end{align}	
with $\mb{p}_{j,i} := \mb{p}_{j}-\mb{p}_{i} := (p_{x,j,i};p_{y,j,i};p_{z,j,i})$ defined as the relative position (distance) of sensors $j$ and $i$ (see Fig.~\ref{fig_radar_drone}). We assume fixed sensors and hence the error term $\|\mb{p}_j\|^2 - \|\mb{p}_i\|^2$ is \textit{fixed} and \textit{known}\footnote{In this paper, we assume the sensors are fixed and static. Therefore the bias term $\|\mb{p}_j\|^2 - \|\mb{p}_i\|^2$ is calculated only once. In case, the sensors are mobile, this term needs to be calculated at every time-instant $k$ and shared over the sensor network.}. Thus, by adding this known error/bias term $\frac{1}{2}(\|\mb{p}_j\|^2 - \|\mb{p}_i\|^2)$ to Eq.~\eqref{eq_h_simp0}, the TDOA measurement can be modified to get the standard form,
\begin{align}	
\label{eq_tdoa_new}
\mb{y}_i(k) = H_i\mb{x}(k) + \bm \nu_i(k),
\end{align}	
Note that this is because the sensors are stationary and both positions $\mb{p}_j$ and $\mb{p}_i$ are known by any sensor $i$. Our model, similar to \cite{usman_loctsp:08,mohammadi2014distributed,ennasr2016distributed,ennasr2020time}, follows zero-mean noise term $\nu_{i,j}(k)$ and synchronized clocks. We refer interested readers to \cite{gong2018auv} for a more detailed analysis on non-zero-mean noise-corrupted measurements.
We use the standard form~\eqref{eq_h_simp}-\eqref{eq_tdoa_new} as our measurement model in the rest of the paper.
Note that for static sensors, the above gives a \textit{constant} output matrix $H_i$, and thus, a standard LTI measurement model.
This is in contrast to the state-dependent (and time-variant) $\overline{H}_i$ from the nonlinear measurement model \eqref{eq_tdoa}. Note that the linearized measurement model in \cite{ennasr2016distributed,ennasr2020time} is given for mobile sensors (UAVs). The results of this work can similarly be extended to that case, for example, by using the fixed (rigid) formation of UAVs via, e.g., the results of \cite{taes,fathian2019robust,ennasr2020time}. However, as explained in the following remark, any variation in the relative distances of UAVs or mobile sensors (due to loose formation setups) may increase the tracking error and degrade the estimation performance.
\begin{rem}
For ``distributed" estimation, the measurements and system model are mainly considered linear in the literature (see, for example, the mentioned STS and DTS methods in the introduction). Similarly, for \textit{distributed} target tracking the nonlinear measurement models need to be linearized, e.g., \cite{ennasr2020time,ennasr2016distributed} use the linearized model \eqref{eq_tdoa}-\eqref{eq_hij} with \textit{time-varying} output matrix $\overline{H}_i$ (which is a function of target's position $\mb{p}$). However, such linearization in $\overline{H}_i$ is naturally subject to certain uncertainty that accentuates the observation error and, in turn, degrades the performance of the distributed estimation. In contrast, our proposed measurement model \eqref{eq_h_simp}-\eqref{eq_tdoa_new} is primarily linear and, further, matrix $H_i$ is independent of $\mb{p}$. Thus, $H_i$ is not prone to positioning and linearization uncertainty. We compare the mean-square-estimation-error (MSEE) performance of both linear and linearized models (based on both estimated and true target positions) via KF in Section~\ref{sec_sim}. The linearization gets worse as the target moves more distant from the sensors, i.e., as $\mb{p}-\mb{p}_i$ increases, which is addressed in \cite{ennasr2020time} by considering a formation of mobile sensors (drones/UAVs) following the target.
The time-varying nature of $\overline{H}_i$ is more problematic for the LMI design of the local feedback gain of the distributed observer. Since the output matrix $\overline{H}_i$ is changing and is a function of the target position $\mb{p}$, the estimate $\widehat{\mb{x}}_i$ (more accurately the first three position terms  in
$\widehat{\mb{x}}_i$) is used instead in $\overline{H}_i$ based on which the gain matrix $K$ needs to be redesigned at every time-step $k$. This iterative recalculation of the gain matrix $K$ adds more complexity and computational load on the sensors. Further, the use of estimate $\widehat{\mb{p}}$ instead of $\mb{p}$ adds more uncertainty in the calculations (particularly for highly noisy setups). 
\end{rem}

\section{Single Time-Scale Networked Estimation} \label{sec_dist}
\subsection{The proposed estimation protocol}
We apply an updated version of the STS networked estimation protocol in \cite{jstsp14,acc13}. The proposed protocol here differs in the sense that no measurement-fusion (\textit{consensus} on the measurement-updates or \textit{innovations} \cite{Li2015Moura,kar2012distributed,das2015distributed}) is needed. In other words, the protocol \textit{only} requires sharing \textit{a-priori estimates} and consensus on \textit{predictions}. This significantly reduces the connectivity requirement of the sensor network.
Considering $W$ as the consensus adjacency matrix of the network, the protocol is proposed as follows:

\noindent \textbf{(i) Prediction over the network: consensus-fusion}
\begin{align}\label{eq_p}
\widehat{\mb{x}}_i(k|k-1) = \sum_{j\in i \cup \mathcal{N}_i} W_{ij}F\widehat{\mb{x}}_j(k-1),
\end{align}

\noindent \textbf{(ii) Local measurement-update (innovation): no fusion}
\begin{align}\label{eq_m}
\widehat{\mb{x}}_i(k) = \widehat{\mb{x}}_i(k|k-1) + K_{i} H_i^\top \left(\mb{y}_i(k)-H_i\widehat{\mb{x}}_i(k|k-1)\right),
\end{align}
where the measurement-update in \eqref{eq_m} is only based on $H_i$ (in contrast to \cite{jstsp14,acc13}). Note that $F\widehat{\mb{x}}_j(k-1)$ represents the current prediction of the target state $\mb{x}$ via sensor $j$ at time $k$, $\widehat{\mb{x}}_i(k|k-1)$ gives the weighted average of these predictions (at time $k$) given all the information of sensor~$i$ and its neighboring sensors~$j \in \mathcal{N}_i$ up-to time $k-1$ as \emph{prior estimate}, $\widehat{\mb{x}}_i(k) \equiv \widehat{\mb{x}}_i(k|k)$  is the updated estimate at time $k$ (using the measurement at time $k$) as \emph{posterior estimate}, and $K_{i}$ is the local observer gain matrix at sensor $i$. $0<W_{ij}<1$ represents the entry (fusion weight at incoming information from sensor $j$) of the \textit{row-stochastic} matrix $W$. The structure (zero-nonzero pattern) of the fusion matrix $W$ as the weighted adjacency matrix of the network follows its graph topology, and the row-stochasticity $\sum_{j\in i \cup \mathcal{N}_i} W_{ij}=1$ follows the consensus nature of the networked estimator. Note that, as compared to \cite{jstsp14},  Eq.~\eqref{eq_m} includes no measurement sharing. This reduces the sensor network connectivity and computational load on sensors as there is no need for direct linking from certain measurement hubs to all other sensors (and no consensus on those measurements), and, as discussed later in Section~\ref{sec_delay}, allows for \textit{delay-tolerant} distributed estimation in the presence of \textit{heterogeneous} time-delays over the network.
\begin{lem} \label{lem_error}
Let $\mb{e}_{i}(k)=\mb{x}(k)-\widehat{\mb{x}}_i(k)$ denote the tracking error at sensor $i$ and $\mb{e}(k)=(\mb{e}_{1}(k); \dots;\mb{e}_{n}(k))$ as the collective error vector. The error dynamics of the networked estimator \eqref{eq_p}-\eqref{eq_m} is,
\begin{align}\label{eq_err1}
	\mb{e}(k) &= (W\otimes F  - K D_H(W\otimes F))\mb{e}(k-1)+
	\bm \eta(k) \\
	&=: \widehat{F} \mb{e}(k-1)+
	\bm \eta(k),
\end{align}
where $\otimes$ denotes the Kronecker matrix product, $D_H := \mbox{blkdiag}[H_i^\top H_i]$ and $K:= \mbox{blkdiag}[K_{i}]$ are defined as block-diagonal matrices with diagonals $H_i^\top H_i$ and $K_{i}$, respectively,
and $\bm \eta(k)=(\bm \eta_{1}(k); \dots;\bm \eta_{n}(k))$ collects the noise terms and random inputs as,
\begin{align} \nonumber
	\bm \eta(k) &= \mb{1}_n \otimes G\mb{w}(k-1) - K\big(D_H(\mb{1}_n \otimes G\mb{w}(k-1)) \\
	&+ \overline{D}_H\bm \nu(k)\big), \label{eq_eta}
\end{align}
with $\bm \nu=(\bm \nu_{1}; \dots;\bm \nu_{n})$,  $\mb{1}_n$ as the column-vector of ones of size $n$, and
$\overline{D}_H \triangleq \mbox{blkdiag}[ H_i]$ as the block-diagonal matrix of $H_i$ (not necessarily a square matrix).	
\end{lem}
\begin{proof}
See the proof in the Appendix.
\end{proof}
\begin{lem} \label{lem_obsrv}
The error dynamics \eqref{eq_err1} can be made stable,
if the pair $(W\otimes F,D_H)$ is observable\footnote{Note that the NCV model has no stable unobservable subspace and the weaker notion of detectability is not the case for the model in this work.}. 	
\end{lem}
\begin{proof}
See the proof in the Appendix.
\end{proof}
Lemma~\ref{lem_obsrv} implies that  $(W\otimes F,D_H)$-observability (referred to as \textit{networked observability} or \textit{distributed observability} \cite{jstsp14}) ensures that matrix $K$ can be designed to stabilize the error dynamics \eqref{eq_err1}. In other words, there exists $K$ such that $\rho (W \otimes F -K D_H (W\otimes F))<1$, where $\rho (\cdot)$ denotes the spectral radius of its argument. Recall that, due to the distributed nature of the protocol, the gain matrix $K$ must be \textit{block-diagonal}.
Such a block-diagonal matrix can be designed, for example, using the local gain design in \cite{khan2011coordinated,rami:97}. This can be done either once offline and embedded to the sensors beforehand or iteratively along with the distributed estimation procedure \cite{khan2011coordinated}. For example, one way to design the matrix $K$ is via iterative cone-complementary optimization
\cite{rami:97} to solve the following LMI,
\begin{align} \label{eq_min}
\begin{aligned}
	\displaystyle
	\min
	~~ &  \mathbf{trace}(XY) \\
	\text{s.t.}  ~~& X,Y\succ 0, ~ & K\mbox{~is~block-diagonal}.\\ ~ & \left( \begin{array}{cc} X&\widehat{F}^\top\\ \widehat{F}&Y\\ \end{array} \right) \succ 0,~& \left( \begin{array}{cc} X&I\\ I&Y\\ \end{array} \right) \succ 0,\\
\end{aligned}
\end{align}
The stopping criteria of the above iterative LMI optimization is $\mathbf{trace}(Y_tX + X_tY)<2Nn + \epsilon$ with a predefined small $\epsilon$ and $N$ as the size of the $F$ matrix which is either $6$ or $9$. We solve this iterative algorithm at the same time-scale $k$ of the target dynamics. This $K$ matrix can be found once offline and, then, the block matrices $K_i$ are given to the static sensors before the tracking process. This is because the sensors are static and the measurement matrix is independent of the target's position. Given that the estimator \eqref{eq_p}-\eqref{eq_m} is observable	(in distributed sense) from Lemma~\ref{lem_obsrv}, the following Lemma states that such a stabilizing block-diogonal $K$ exists.
\begin{lem} \label{lem_usman}
\cite[Lemma~2]{khan2011coordinated} If the pair $(W\otimes F,D_H)$ is generically observable, then a (structured) block-diagonal gain matrix, $K$, is the
solution of the optimization~\eqref{eq_min}. 	
\end{lem}

\begin{rem}
For the time-varying measurement matrix $\overline{H}_i$ given by Eq. \eqref{eq_tdoa}-\eqref{eq_hij} (as considered by \cite{ennasr2020time}), the gain matrix $K$ needs to be updated at every time-step $k$  based on
$\overline{H}_i$. This approach considerably increases the algorithm complexity and computational load on the sensors, particularly in real-time. However, for the proposed constant measurement matrix $H_i$, $K$ is only designed once offline and embedded to all sensors with no need to be redesigned in real-time tracking, which makes our solution superior to the existing distributed estimation methods using linearized $\overline{H}_i$, e.g., \cite{ennasr2016distributed,ennasr2020time}. See Section~\ref{sec_sim_lin} for more detailed illustration.
\end{rem}

Following Lemma~\ref{lem_obsrv}, for the pair $(W\otimes F,D_H)$ to be observable, the structure of $W$ (i.e., the sensor network connectivity)  needs to satisfy some properties as discussed next.

\subsection{Structural  design of the sensor network} \label{sec_dist_net}
Recall that, from structured systems theory \cite{woude:03,khan2011coordinated}, observability in Lemma~\ref{lem_obsrv} depends on the \textit{structure} of $W\otimes F$ and $D_H$, irrespective of their exact numerical values, i.e., if $(W\otimes F,D_H)$-observability holds for one choice of $W\otimes F$ and $D_H$, it holds for \textit{almost all} numerical values of these matrices holding the same structure.
Recall from Eq.~\eqref{eq_p} that the structure of $W$ is tied with the topology of the sensor network $\mc{G}_W$, over which the sensors share their position information, TOA values, and their estimates.
The observability analysis is tightly related to the structural or generic rank (or \textit{G-rank}) of the transition matrix $F$. Define the G-rank as the maximum matrix rank that can be obtained by reassigning the numerical entry values while keeping the matrix structure fixed\footnote{For a system matrix, the G-rank can be determined by the size of the \textit{maximum matching} in its associated \textit{system digraph},
	which is equal to the maximum number of matrix elements no two of which share rows/columns \cite{murota}. }.

\begin{lem} \label{lem_rank}
	The transition matrix $F$, given by \eqref{F_ncv} and \eqref{F_nca}, is full G-rank.
\end{lem}
\begin{proof}
	See the proof in the Appendix.
\end{proof}
In a system digraph, define the parent SCC (or parent node)
\cite{jstsp14} (also referred to as the root SCC/node \cite{pequito2015framework}) as a SCC/node with no outgoing links to other SCCs/nodes. The term SCC stands for Strongly-Connected-Component, i.e., a graph component in which there is a directed path from every node to all other nodes in the same component. In terms of its adjacency matrix, any upper (or lower) block-triangular \textit{reducible} matrix can be decomposed into \textit{irreducible}\footnote{An irreducible matrix cannot be transformed into block upper-triangular or block lower-triangular by simultaneous row/column permutations.} block diagonals where each of them represents an SCC in adjacency graph visualization; the diagonal blocks with no lower (resp. upper) diagonal entries, represent an SCC with no outgoing links to other SCCs, i.e., to other block diagonals. See \cite{algorithm} for more details.  Then, the following holds.
\begin{lem}
	For a full G-rank system matrix $F$, the necessary and sufficient condition for (structural) observability is achieved by measuring every parent SCC/node in the system digraph $\mc{G}_F$. \label{lem_parent}
\end{lem}
\begin{proof}
	See the proof in the Appendix.
\end{proof}
For the system matrices \eqref{F_ncv} and \eqref{F_nca}, the parent nodes in the system digraph are associated with  $p_x,p_y,p_z$.


\begin{thm} \label{thm_sc}
	Given the linear measurement model~\eqref{eq_h_simp}-\eqref{eq_tdoa_new}, the pair $(W\otimes F,D_H)$ is observable if the network $\mc{G}_W$ is strongly-connected (SC).
\end{thm}
\begin{proof}
	See the proof in the Appendix.
\end{proof}
It is typical in target tracking scenarios to assume bidirectional communications and data-sharing among the sensors \cite{ennasr2020time}, implying mutual information exchange (message-passing) between every two neighbors $i$ and $j$. In this case,  connectivity of  $\mc{G}_W$
ensures distributed observability in Theorem~\ref{thm_sc}. Such networks can be optimally designed via the results in \cite{tnse19} in polynomial-order complexity.
Note that, since the information of each sensor is shared with other sensors via a path (a sequence of communicating sensors), the prior estimates (or predictions) of every sensor $i$ (on the target's state) would eventually reach and get updated at every other sensor at most in $d_n$ time-steps (with $d_n$ as the network diameter). A similar structural observability analysis is given in \cite{ennasr2020time}.

\subsection{Tracking in Presence of Heterogeneous Delays } \label{sec_delay}
In this section, we go one step further and extend the results to provide tracking stability in the presence of heterogeneous delays on different links over the sensor network.
\begin{ass}
	The assumptions on the delays are as follows:
	\begin{itemize}
		\item The delays are fixed and heterogeneous. We assume $\tau_{ij}$ is \textit{constant} for every link $(j,i)$ over the static sensor network. The delay $\tau_{ij}$s are heterogeneous for different links $(j,i)$, in general.
		\item The transmitted data packets are \textit{time-stamped}, so the time-delay $\tau_{ij}$ is \textit{known} from sensor $j$ at the receiver sensor $i \leftarrow j$.
		\item The max possible delay at all links is $\overline{\tau}:=\max_{i,j} \{ \tau_{ij} \}$ and is assumed to be finite. The latter ensures that there is no packet loss and missing data over the sensor network.
	\end{itemize}
\end{ass}
The protocol \eqref{eq_p}-\eqref{eq_m}, then, changes into:
\begin{align} \nonumber
	\widehat{\mb{x}}_i(k|k-1) =& W_{ii}F\widehat{\mb{x}}_i(k-1) \\\label{eq_p2} &+ \sum_{j\in\mathcal{N}_i} \sum_{r=0}^{\overline{\tau}} W_{ij}F^{r+1}\widehat{\mb{x}}_j(k-r) \mb{I}_{k-r,ij}(r),
	\\\label{eq_m2}
	\widehat{\mb{x}}_i(k) =& \widehat{\mb{x}}_i(k|k-1) + K_i H_i^{\top} \left(\mb{y}_i(k)-H_i \widehat{\mb{x}}_i(k|k-1)\right),
\end{align}
Note that there is no information-sharing involved in Eq.~\eqref{eq_m2}, and delays only affect Eq.~\eqref{eq_p2} where over every link $(j,i)$, $\mb{I}_{k-r,ij}(r)$ is the indicator function defined as,
\begin{align}
	\mb{I}_{k,ij}(r) = \left\{
	\begin{array}{ll}
		1, & \text{if}~ \tau_{ij}=r  \\
		0, & \text{otherwise}.
	\end{array}\right.
\end{align}
which is non-zero \textit{only} for $r=\tau_{ij} \in [0~\overline{\tau}]$. This follows from our constant delay assumption.
\begin{thm} \label{thm_tau}
	Let $(W\otimes F, \overline{D}_H) $ be observable and assume  block-diagonal feedback gain matrix $K$ is given for the delay-free case such that $\rho(W\otimes F  - K D_H(W\otimes F))<1$; then, for $\overline{\tau}$ satisfying,
	\begin{align} \label{eq_tau*}
		\rho(W\otimes F^{\overline{\tau}+1} - K \overline{D}_H (W\otimes F^{\overline{\tau}+1}))< 1.
	\end{align}
	the error dynamics of Eq.~\eqref{eq_p2}-\eqref{eq_m2} remains Schur stable.
\end{thm}
\begin{proof}
	See the proof in the Appendix.
\end{proof}
Inequality~\eqref{eq_tau*} implies Schur stability of the error dynamics under possible heterogeneous fixed delays at different links. The same $K$ matrix designed for the delay-free case can be used in the delayed case, which makes the algorithm delay-tolerant with \textit{no need to redesign} $K$. For the same $K$ matrix, the term in the LHS of \eqref{eq_tau*} becomes closer to $1$ as the $\overline{\tau}$ (the max delay bound) increases. As the spectral radius of the error dynamics increases, the error convergence rate decreases.
Although we assume fixed delays, for monotone systems one can assume these fixed values as the maximum possible transmission delays at every link \cite{charalambous2015distributed}.

\section{More Discussions and Auxiliary Results} \label{sec_auxiliary}
\subsection{$q$-Redundant distributed observer design}
One can design distributed observers/estimators resilient to link (and node) removal over the sensor network, referred to as $q$-redundant networked observers. Recall that from Theorem~\ref{thm_sc}, the sensor network only needs to be SC for distributed observability. There exist efficient algorithms to design $q$-edge-connected (or $q$-node-connected) networks that remain SC after the removal of up-to $q$ links (or $q$ nodes). This is also called \textit{survivable network design} \cite{umsonst2019tuning,jabal2021approximation}.
Over such sensor networks, the distributed estimator \eqref{eq_p}-\eqref{eq_m} can tolerate up to $q$ missed links or failed sensors, while the target remains observable (in the distributed sense) to all the remaining sensors over the (possibly reduced) sensor network. However, the gain matrix $K$ needs to be redesigned (e.g., in an iterative procedure over the time $k$ \cite{khan2011coordinated}). The simulations in the next section better illustrate this.

\subsection{Brief overview on adding local fault detectors}

Some of our results on  \textit{joint distributed observer and FDI} in case the TOA measurement received by sensor $i$ is faulty (or largely biased) are reviewed here. These measurement biases could be due to sensor localisation errors or potential errors in the translation of the time
of reception (TOAs) into relative distance (range). Assuming Gaussian process noise $\mb{w}(k)$ and measurement noise $\nu_i(k)$, the results follow our previous work in \cite{tcns2021}. Considering a fault term, the measurement model takes the form
$\mb{y}_i(k) = H_i\mb{x}(k) + \bm \nu_i(k) + \mb{f}_i(k)$
with the \textit{additive fault} term $\mb{f}_i(k)$. This is reflected in the error dynamics and the noise term in \eqref{eq_eta} changes to,
\begin{align} \nonumber
	\bm \eta (k) &= \mb{1}_n \otimes G\mb{w}(k-1) - K\big(D_H(\mb{1}_n \otimes G\mb{w}(k-1)) \\
	&+ \overline{D}_H\bm \nu(k)\big) - K \overline{D}_H \mb{f}(k), \label{eq_eta_f}
\end{align}
with $\mb{f}=(\mb{f}_1;\dots;\mb{f}_n)$ as the column concatenation of all fault terms. Since both $K$ and $\overline{D}_H$ are defined block diagonal, the fault term $\mb{f}_i$ at every sensor $i$ \textit{only} appears in $\eta_i$ and is isolated from the rest of the network. This differs with \cite{tnse2021} where a specific LMI gain design needs to be adopted to isolate fault (or attack) at every sensor. Every sensor $i$ finds the absolute residual $\mb{r}_i(k)=|\mb{y}_i(k)-H_i\widehat{x}_i(k)|$ and detects possible faults via certain probabilistic thresholds based on some confidence-intervals on the fault-free residuals, see details in \cite[Fig.~3]{tnse2021}.
For a given FAR (false-alarm-rate) $\varkappa$ and its threshold, $\mc{T}_\varkappa$, as a function of $\varkappa$, and $\Phi:=\mathbb{E}(\bm \eta^\top_k \bm \eta_k)$ assuming Gaussian distribution for noise and input variables (see \cite{khan2014collaborative,tcns2021} for details), the sensor triggers the alarm whenever $\mb{r}_i(k) \geq \mc{T}_\varkappa$.
One can also take the history of the residuals over a sliding time-window $\theta_{\chi}$ (stateful detection), e.g., via \textit{distance measures} (instead of  instantaneous residuals $\mb{r}_i(k)$) as \cite{icas21_attack},
\begin{align} \label{eq_z}
	\mb{z}_i(k) = \sum_{m=k-\theta_{\chi}+1}^k \frac{(\mb{r}_{i}(k))^2}{\Phi},~k \geq \theta_{\chi},
\end{align}
It can be shown that this measure follows the $\chi^2$ (Chi-square) distribution assuming i.i.d. random variables $\mb{r}_{i}$'s, see details in \cite{chi_book}. Then, the associated probabilistic threshold (equivalent to $\mc{T}_\varkappa$) on this measure for FAR $\varkappa$ is defined based on $\chi^2$ distribution.
Recall that the above detection techniques are distributed (or localized) at sensors with no need of a centralized coordinator or decision-maker, i.e., each sensor locally detects any fault at its measurement based on its available local data.

\subsection{Some relevant remarks}
The following remarks are noteworthy:
\begin{itemize}
	\item Clock synchronization algorithms \cite{a8030590,maroti2004flooding,4434671}  can be used to synchronize the clocks. The assumption that sensors/agents are synchronized is common in distributed sensor networks and consensus literature, e.g., \cite{ennasr2016distributed,ennasr2020time,he2020secure,olfati2011collaborative}.
	\item  The STS networked estimation \eqref{eq_p}-\eqref{eq_m} performs one-step of averaging (or consensus fusion) between two successive time-steps $k$ and $k+1$ of the target dynamics \eqref{eq_targ}, in contrast to the DTS scenarios in \cite{ olfati2011collaborative, battistelli2014consensus,he2020secure,battilotti2021stability}, where many iterations $L\geq d_n$ of information-exchange and consensus on a-priori estimates are implemented at a time-scale much faster than $k$. This requires a much faster communication and computation rate than the sampling rate of the target dynamics (i.e., $k$ in \eqref{eq_p}-\eqref{eq_m}).
	In STS estimation, many works \cite{mohammadi2014distributed,Li2015Moura,kar2012distributed,das2015distributed,wu2019efficient,rastgar2018consensus} assume local observability, i.e., all necessary information (e.g., $3$ neighboring measurements \cite{usman_loctsp:08} and predictions) are directly communicated to every sensor from its neighbors. This needs more connectivity.
	However, in general, more data-sharing over the network improves the estimation performance. Similarly, many DTS protocols only need connected undirected networks (requiring a faster communication rate $L \geq d_n$).
	
	\item Following the previous comment, Table~\ref{tab_compare} compares the proposed protocol with some relevant distributed estimation and FDI literature in terms of connectivity and rate of data-exchange. Recall from \cite{giraldo2018survey} that, for SLS models, there is no knowledge of system dynamics and, thus, more measurements and network connectivity are needed in general.  This table shows how our proposed method, similar to \cite{ennasr2020time}, reduces the communication burden on sensors, and therefore, the cost of communication devices in real-time applications. This is in terms of expenses regarding both the rate of data-sharing (e.g., communication speed) and data-transmission traffic (network connectivity). See details in Section~\ref{sec_intro} and  Fig.~\ref{fig_fusion}.
	\begin{table} [t]
		\centering
		\caption{Comparison between different distributed estimation (and detection) methods for SLS and LDS systems in terms of network-connectivity $\times$ communication-rate. }
		\label{tab_compare}
		\begin{tabular}{|c|c|c|c|}
			\hline
			Ref. & time-scale & system dynamics & links $\times$ rate    \\
			\hline
			\cite{sauter:09} & STS & SLS &  $N(N-1)\times 1$   \\
			\hline
			\cite{Li2015Moura,kar2012distributed,das2015distributed} & STS & LDS &  $3N\times 1$  \\\hline
			This work	 & STS  & LDS &  $N\times 1$  \\\hline
			\cite{he2020secure,olfati2011collaborative} &	DTS & SLS/LDS &  $N\times L$ with $L\geq d_n$\\
			\hline
			\hline
		\end{tabular}
	\end{table}
	\item  Note that, in the case of \textit{mobile} sensor networks, the constant delay assumption may not necessarily hold; this makes the latency analysis more challenging as the network topology and the distances change. Formation setups are needed to fix the distance in the case of mobile sensors \cite{fathian2019robust,ennasr2020time,taes}. However, such simultaneous decentralized swarm formation plus target tracking scenarios \cite{ennasr2020time}  may not necessarily work in the presence of heterogeneous delays as considered in the current paper.
	\item As stated in Theorems~\ref{thm_sc} and \ref{thm_tau}, strong-connectivity of the sensor network is sufficient for distributed observability of the target and Schur stability of the error dynamics. Having more than strong-connectivity implies more information sharing over the network and more efficient noise filtering. Further, with more connectivity, the gain design procedure \eqref{eq_min} reaches stabilizable $K$ matrix via fewer iterations. On the other hand, this increases the network traffic and communication costs. 
	
\end{itemize}

\section{Simulation Results} \label{sec_sim}
\subsection{Linear versus linearized TDOA models: centralized case} \label{sec_sim_lin}
\begin{figure} [t]
	\centering
	\includegraphics[width=1.75in]{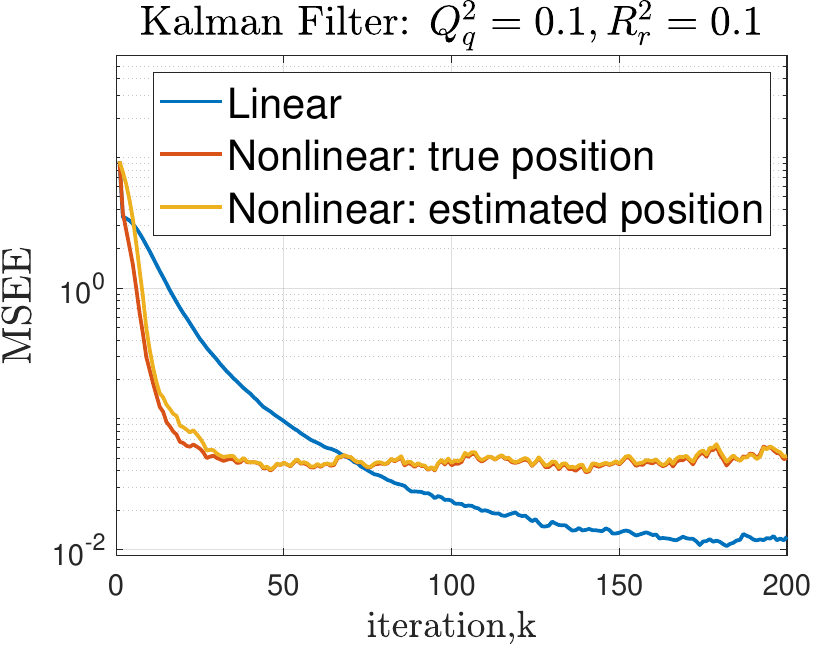}
	\includegraphics[width=1.75in]{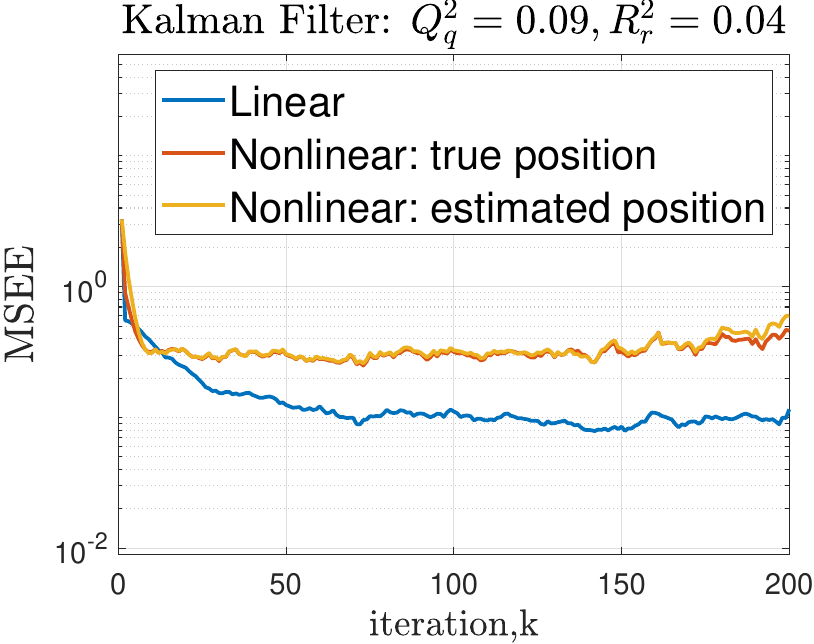}
	\includegraphics[width=1.75in]{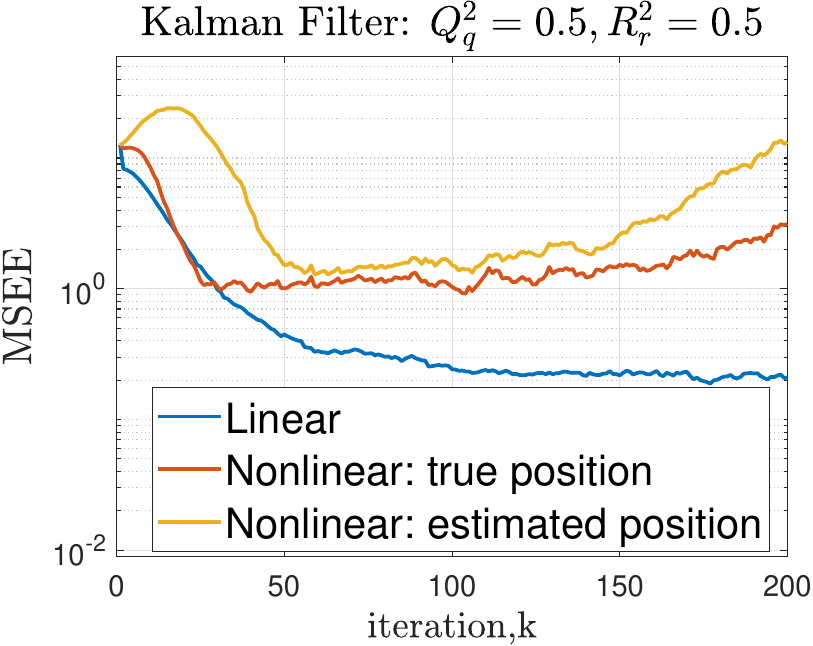}
	\includegraphics[width=1.75in]{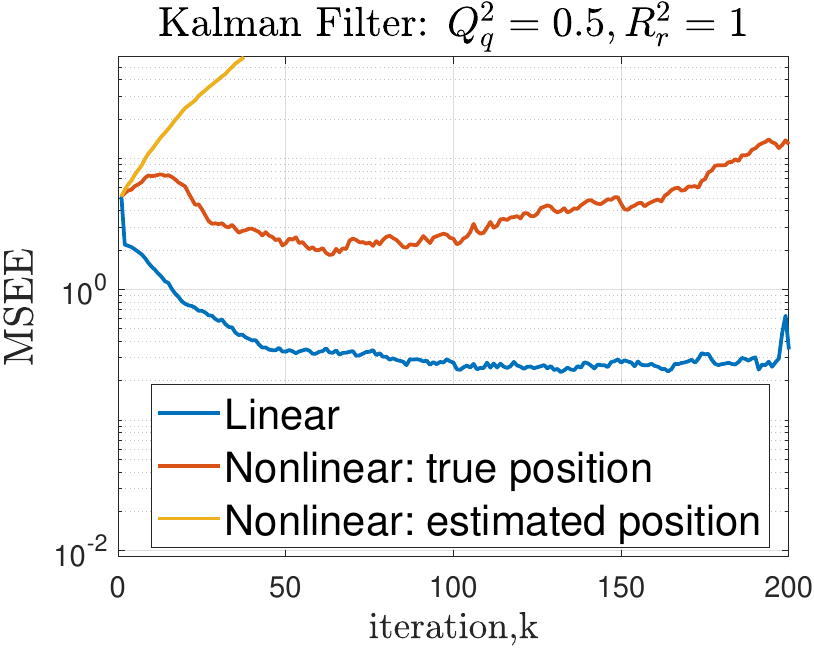}
	\caption{This figure compares the MSEE performance of the linear measurement model \eqref{eq_h_simp}-\eqref{eq_tdoa_new} versus the linearized model \eqref{eq_tdoa}-\eqref{eq_hij} used in \cite{ennasr2016distributed,ennasr2020time} for different measurement and process noise variances. Note that, for the nonlinear case, the linearized output matrix $\overline{H}_i$ is a function of the target position itself. For the simulation, both the true target position and estimated target position are considered in the output matrix for the sake of simulation.  }
	\label{fig_kf}
\end{figure}

In this subsection, we compare the MSEE performance of our proposed linear measurement model  \eqref{eq_h_simp}-\eqref{eq_tdoa_new} versus the linearization of the nonlinear model \eqref{eq_tdoa}-\eqref{eq_hij} (used in \cite{ennasr2020time,ennasr2016distributed}) via \textit{centralized} KF.  For this purpose, we perform Monte-Carlo (MC) simulation (averaged over $100$ trials) on the NCA target model \eqref{F_nca} with parameters: $ T=0.004$, $n=7$, $\nu_{i,j} = \mc{N}(0,R_r^2)$, $\mb{w} = \mc{N}(0,Q_q^2\mb{I}_6)$ for different noise variances $Q_q,R_r$ as stated in Fig.~\ref{fig_kf}\footnote{Our general assumption on the noise is zero-mean white noise. Gaussian noise in the simulation is only considered as a typical example.}. The position of sensors $\mb{p}_i$ and initial target position $\mb{p}(0)$ are randomly chosen in the range $[0,10]$ for $X,Y,Z$-coordinates. Recall that in the linearized measurement model \eqref{eq_y_lin}-\eqref{eq_hij}, the output matrix $H_i$ is a function of the target position $\mb{p}$. The received nonlinear TDOA measurement \eqref{eq_tdoa} is based on the true position of the target $\mb{p}$; however, for the distributed scenario (as in \cite{ennasr2016distributed,ennasr2020time}), for the LMI design of the gain matrix $K(k)$ based on the system and output matrices $F,\overline{H}(k)$, the estimated position of the target based on $\widehat{x}(k)$ is considered (since the exact position $\mb{p}$ is unknown).
Comparing the mean squared estimation error (MSEE) for three models in Fig. \eqref{fig_kf}, the linear case shows better steady-state performance compared to the linearized case, particularly for highly noisy systems.

\subsection{Simulating the distributed NCV tracking} \label{sec_sem_ncv}
\begin{figure*} 
	\centering
	\includegraphics[width=1.75in]{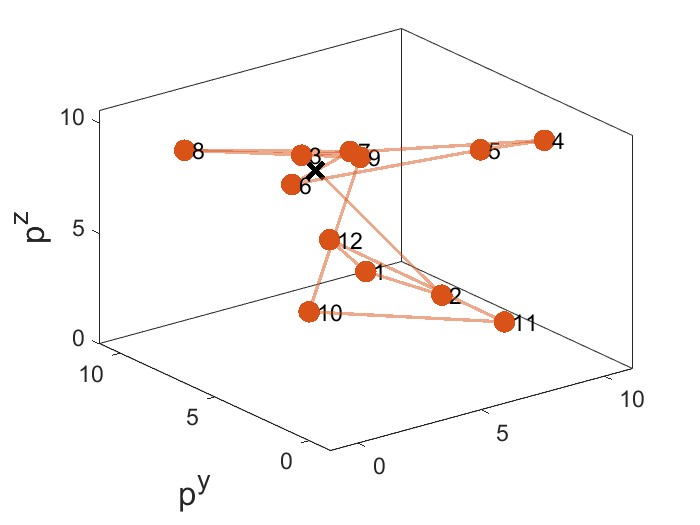}
	\includegraphics[width=1.75in]{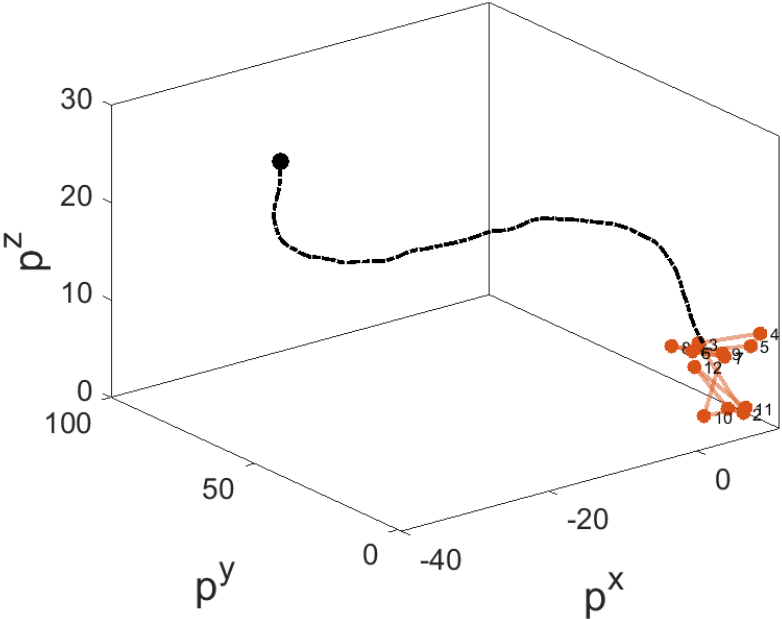}
	\includegraphics[width=1.75in]{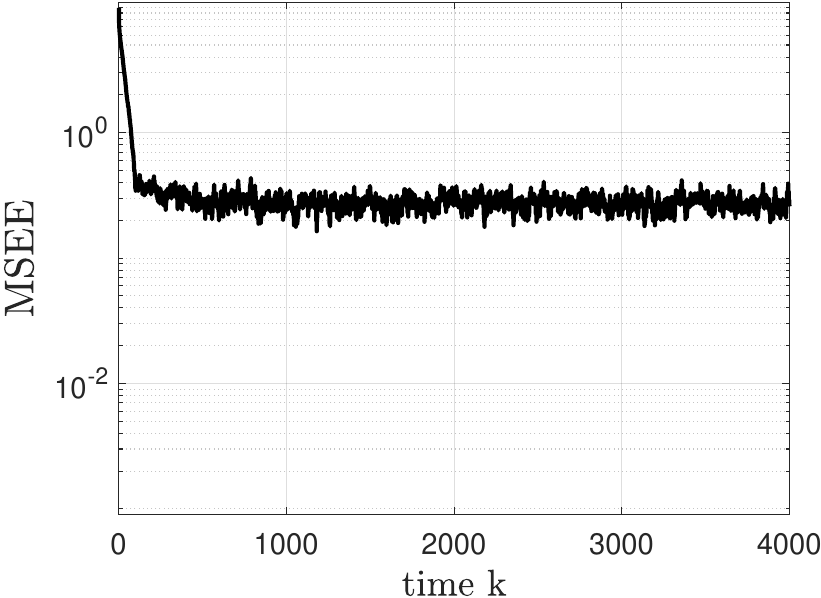}
	\includegraphics[width=1.75in]{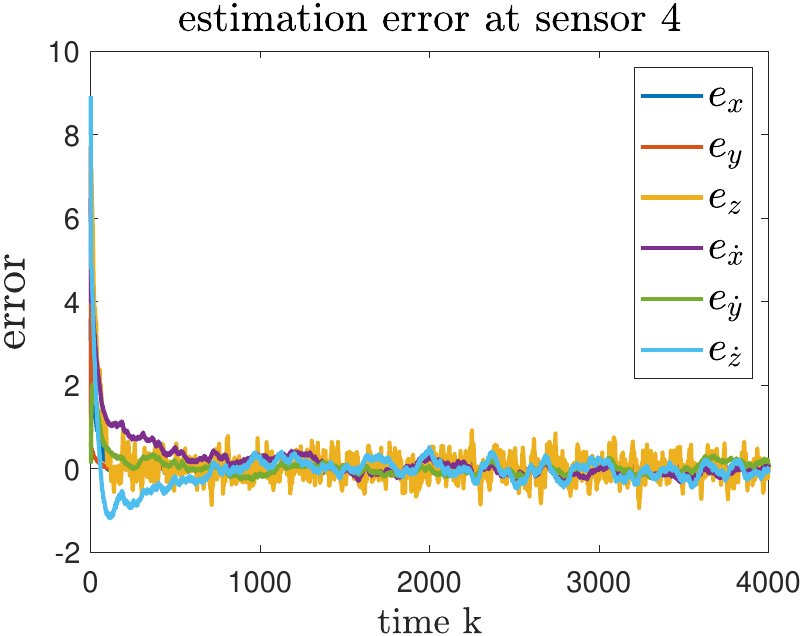}
	\includegraphics[width=1.75in]{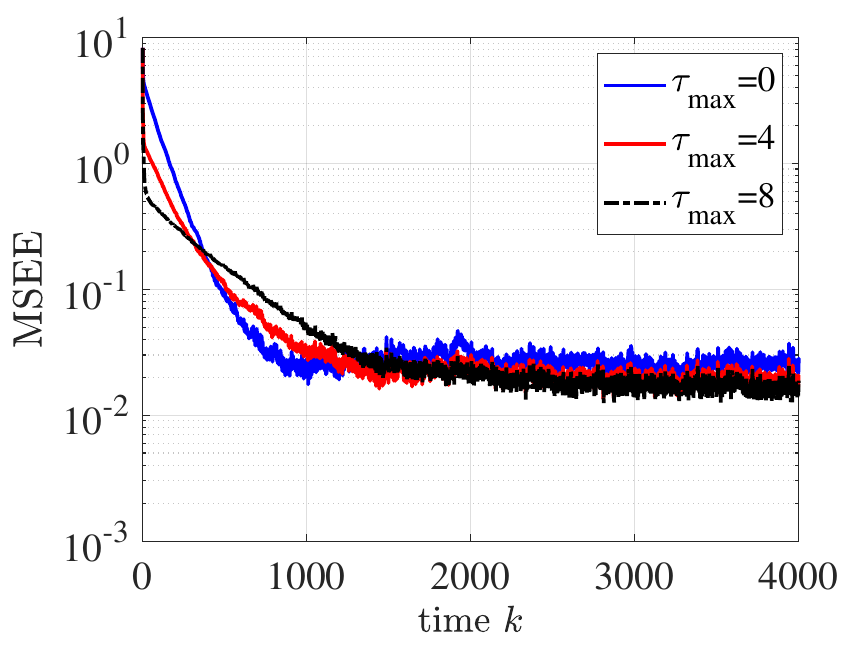}
	\includegraphics[width=1.75in]{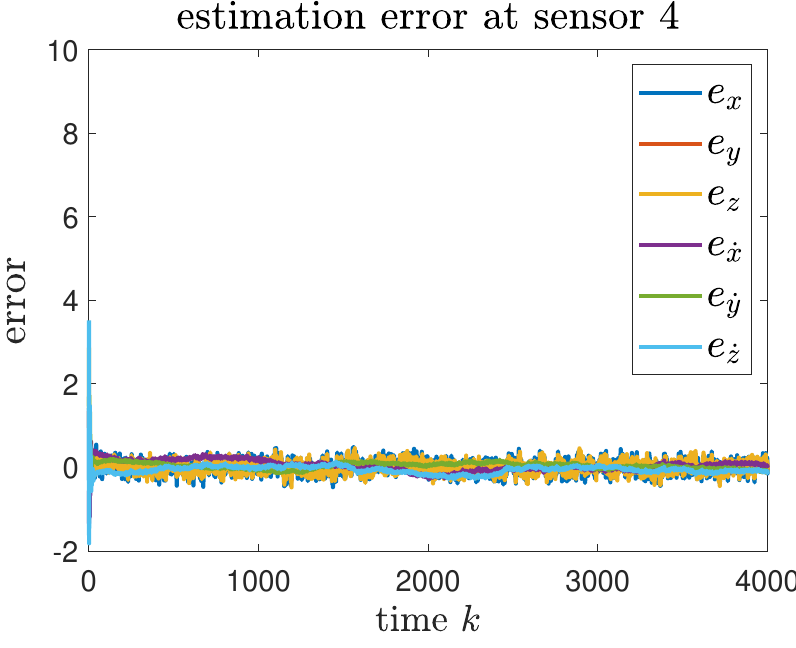}
	\includegraphics[width=1.75in]{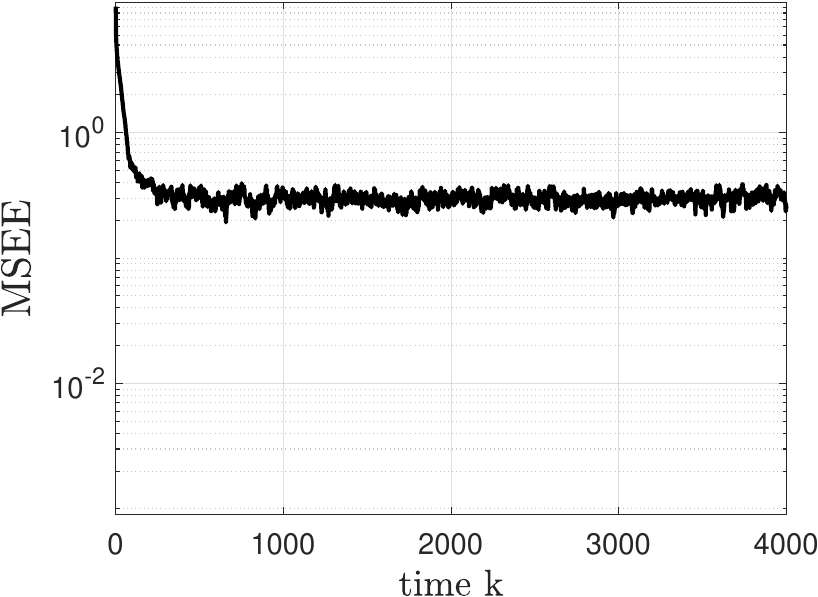}
	\includegraphics[width=1.75in]{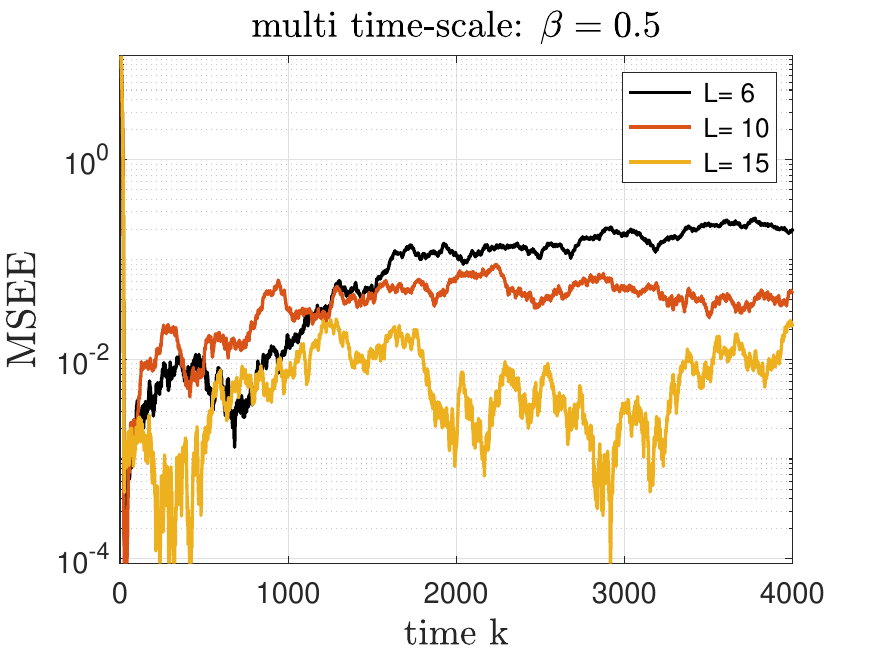}
	\includegraphics[width=1.75in]{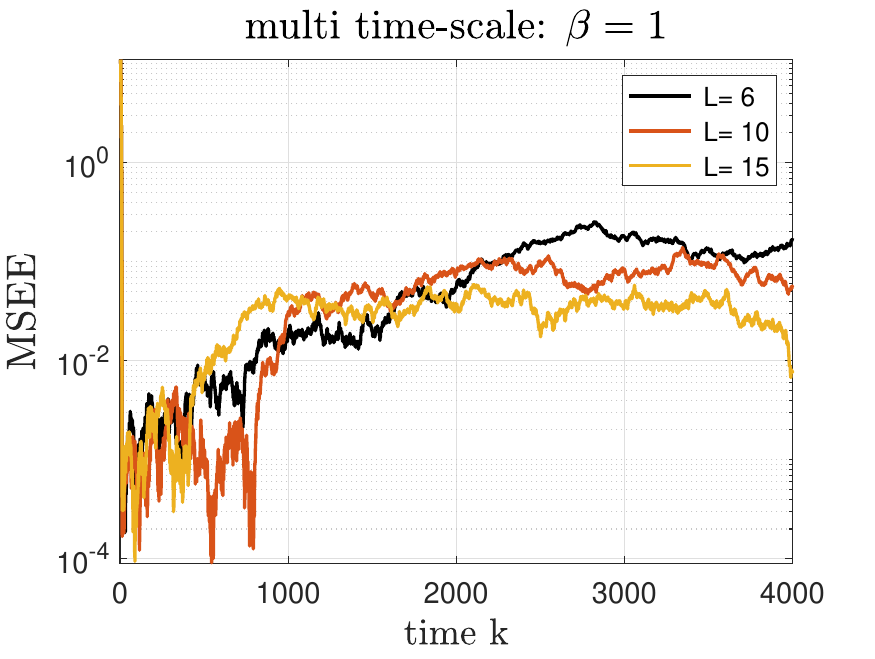}
	\caption{This figure shows the performance of the STS protocol \eqref{eq_p}-\eqref{eq_m} for tracking the NCV target compared to the DTS protocol in \cite{he2020secure}. (LeftTop) The position of sensors, their cyclic communication network (red-colored), and the initial position of the target (black cross), (MidTop) the target path and final position (black-colored) are shown over $4000$ time-steps. (RightTop) The MSEE performance averaged at all sensors over $20$ MC trials is shown, which is steady-state stable and implies observable estimation. (RightTop) The position and velocity errors at (an example) sensor $4$ are unbiased in steady-state. (LeftMid)  Target tracking subject to heterogeneous delays at different links with max delays $\overline{\tau} = 4,8$:  MSEE MC simulation averaged over $10$ trials for all sensor states and (MidMid) state errors at a randomly chosen sensor with $\overline{\tau} = 4$. The noise variances are different for this example. (RightMid) The (averaged) MSEE over the reduced sensor-network after the removal of one link is steady-state stable and thus distributed observability is preserved after link failure. (Bottom) The MSEE performance of the DTS protocol with $L$ additional epochs of averaging (consensus) and communication between every two consecutive time-steps for two different $\beta$ values are presented. For this DTS simulation, more measurements are considered at every sensor (since the system is marginally stable \cite{he2020secure}) such that the target is locally observable. From the figure, as the target moves away from the sensors the MSEE performance degrades and more consensus epochs (larger $L$) might be needed. }
	\label{fig_ncv}
\end{figure*}

For simulation, the NCV model is considered with  $\mb{w}(k)$ in \eqref{eq_targ} as zero-mean white Gaussian random input. Sensors and initial target are positioned randomly in 3D with $0<p_{x,i},p_{y,i},p_{z,i}<10$ and initial values are considered for the target position and $\widehat{\mb{x}}_i(0)$ as its estimates. Measurement noise is $\bm \nu_i\sim\mc{N}(0,0.04)$, and the simulations for error dynamics are averaged over $20$ Monte-Carlo (MC) trials.
For this simulation, $10$ sensors are considered to communicate over a cyclic network. The sampling time in ~\eqref{eq_targ}-\eqref{F_ncv} is $T = 0.02$. The fusion (consensus) weights in $W$ are randomly selected such that it is row-stochastic with its irreducible structure following the symmetric cyclic network topology (similar to \cite{ennasr2020time}). Each sensor $i$ performs the networked estimation \eqref{eq_p}-\eqref{eq_m} and shares its estimates, TOA values, and position with two neighboring sensors $j \in \mc{N}_i$. Note that the block-diagonal gain $K$ is designed once offline via the LMI in \cite{rami:97} and the gain $K_i$ is provided to the sensor $i$. Recall that the TOA values are used at sensor $i$ to find the TDOA measurement \eqref{eq_h_simp}-\eqref{eq_tdoa_new}.
Fig.~\ref{fig_ncv} shows the target path along with the sensors' positions (and the network). MSEE averaged at all $10$ sensors is given in Fig.~\ref{fig_ncv}-RightTop. Further, we showed the error performance at sensor $4$ in Fig.~\ref{fig_ncv}-LeftMiddle as an example. Evidently, the tracking error of all states (positions and velocities) is stable, implying that the sensor can localize the target over time with bounded steady-state error. Next simulations are devoted to tracking in the presence of heterogeneous delays, where $\tau_{ij}$'s are randomly chosen (using MATLAB \textsc{rand} command) in the range $0$ to $\overline{\tau}$. The simulation results are shown in the middle row of Fig.~\ref{fig_ncv}-MidMiddle for max delays $\overline{\tau}=4,8$. It can be easily checked that for these values of $\overline{\tau}$ Theorem~\ref{thm_tau} holds, i.e., for $\overline{\tau}=4,8$  the $\rho(\cdot)$ values in Eq.~\eqref{eq_tau*} are respectively $0.997,0.998$ and are less than $1$, implying that Schur stability holds under the delayed cases. Note that for delay-free case ($\overline{\tau}=0$) we have $\rho(\cdot)= 0.996 <1$.
Note that the MSEE performance of the \textit{centralized} Kalman filter in general might be better than distributed filtering. This is because in the centralized case information of all sensors are available to every sensor (i.e., to consider an all-to-all network). This significantly improves the filtering performance as compared to the sparse connectivity in the distributed case. However, in large-scale, this all-to-all connectivity comes with increasingly costly communications and high network traffic, which is not practical in many real applications.  

\subsection{Redundant distributed observability}
For the example sensor network in Section~\ref{sec_sem_ncv}, the cyclic network is $1$-redundant ($1$-node/link-connected), i.e., by removing any $1$ link (or node) the remaining network is still connected. Following Theorem~\ref{thm_sc}, this implies that the target is distributed observable (in the generic sense) to the remaining sensor network. In this case, assume a random (faulty) communication link between nodes $1$ and $10$ to be removed. The remaining sensor network (a line graph) is $0$-redundant, and the MSEE performance over this reduced network is shown in Fig.~\ref{fig_ncv}-MidRight. The MSEE is bounded steady-state stable, implying that the target is still (distributed) observable over the remaining network. Note that, after link removal, the fusion weights at the neighboring sensors (adjacent to the removed links) need to be re-adjusted such that $W$ remains row-stochastic, e.g., the weight of the removed link $W_{ij}$ and $W_{ji}$ are added to the diagonals $W_{ii}$ and $W_{jj}$ (the self links), respectively. Further, the $K$ matrix needs to be redesigned based on the new $W$ matrix.
Simulation on the same tracking example shows how the proposed $q$-redundant distributed setup tolerates the restructuring (e.g., due to faults).

\subsection{Comparison with the DTS methods}
We compare the MSEE performance with the recent DTS method in \cite{he2020secure}. Recall that the STS strategy in \cite{mohammadi2014distributed,Li2015Moura,kar2012distributed,das2015distributed,wu2019efficient} require more connectivity compared to the cyclic network in our simulation, while the DTS methods including \cite{he2020secure}, work over connected networks.
Fig.~\ref{fig_ncv} provides a comparison of the average MSEE performance with the DTS tracking strategy in \cite{he2020secure}, with its parameters chosen as\footnote{The parameter $\beta$ is denoted as the observation confidence in \cite{he2020secure}, which reflects the trade-off
	on the corrupted and true measurement updates. In other words, for small $\beta$ the measurement-update contribute less to the filtering performance (in case the measurements are corrupted and faulty due to attacks), while, on the other hand, large $\beta$ values imply that the measurement-updates contribute more in the filtering (in case the observations are more trustworthy).  $\alpha$ and $L$ are the consensus parameters that must be tuned. $L\geq d_n$ with $d_n$ as the network-diameter ensures that the information of every node reaches every other node between every two iterations (hence the observability follows). Parameter $\alpha$, as discussed in \cite{he2020secure}, is tuned based on the network structure and the link weights. }: $\alpha =1.8$,  $\beta=0.5,1$, $L=6,10,15$.  We performed the simulation for two values of the \textit{observation conﬁdence} parameter $\beta$  to address the trade-off discussed in \cite{he2020secure}: for small $\beta$, the saturation level is lower to reduce the effect of noise on real observations, while the contribution of the measurement updates is less in the estimation performance. This implies more consensus iterations (larger $L$) to improve filter performance as the target moves far from the sensors (see Fig.~\ref{fig_ncv}-BottomLeft). On the other hand, larger $\beta$ puts more value on the real noisy measurements, which may result in more oscillatory behaviour (see Fig.~\ref{fig_ncv}-BottomRight) in the error dynamics (and even error divergence\footnote{As we tried for large saturation levels in \cite[Algorithm~1]{he2020secure}, e.g., greater than $1$, the MSEE performance considerably degrades for this marginally stable system, and the error (probably) diverges as stated in \cite[Section~3.A]{he2020secure}.}). Recall that the parameter $L$ represents the number of consensus, data-processing, and information-exchange epochs between $k$ and $k+1$, see Fig.~\ref{fig_fusion}. Therefore, \cite{he2020secure} needs $L$ times faster communication and processing rates than the proposed estimator in this paper.
Note that due to saturation effects in \cite{he2020secure} (for security concerns), the tracking performance might be degraded as the target's distance from the sensors increases over time. Note that, $L \geq d_n$ with (network-diameter) $d_n=5$ in this network example. It should be noted that this simulation considers more measurements at every sensor such that the target is \textit{observable locally}. This redundancy significantly improves the error performance compared to the proposed protocol in Fig.~\ref{fig_ncv}-Bottom with only \textit{distributed observability}.

\section{Conclusions and Future Directions} \label{sec_con}
This paper considers a delay-tolerant distributed tracking method with LTI measurements with no local observability assumption.
The proposed STS distributed estimation mandates less communication/computation rates on sensors than DTS and similar STS tracking/estimation methods, while the linear measurement model improves the error performance. This is of more interest in real-time applications.
As future research direction, sensor failure recovery and countermeasures to recover for loss of data while addressing structural optimality \cite{spl17},  optimal design of the sensor network subject to distributed observability  \cite{tnse19,pequito2017structurally}, and optimal sensor power allocation for remote state estimation are of interests. Mobile robotic networks and formation scenarios for dynamic target tracking subject to heterogeneous delays are another interesting research direction. Note that, as mentioned in Section~\ref{sec_intro}, for mobile sensors, e.g., a swarm of UAVs under (barycentric-coordinate + distance-based) formation scenario as in \cite{fathian2018distributed}, coordination and reaching formation under heterogeneous delays might be challenging and latency may even destabilize such formation setups. Therefore, those mobile formation scenarios may not necessarily work in the presence of large and/or heterogeneous delays and need further analysis (as future research direction).

\section*{Appendix: Proofs}

\textbf{Proof of Lemma~\ref{lem_error}:}
\begin{proof}
	The estimation error at every sensor $i$ is,
	\begin{align}
		\mb{e}_{i}(k) &=\mb{x}(k) - \sum_{j\in i \cup \mathcal{N}_i} W_{ij}F\widehat{\mb{x}}_j(k-1) \nonumber \\ &- K_{i} H_i^\top\Big(\mb{y}_i(k)-H_i\sum_{j\in i \cup \mathcal{N}_i} W_{ij}F\widehat{\mb{x}}_j(k-1) \Big).  \nonumber
	\end{align}
	Substituting the system equation \eqref{eq_targ} and measurement equation \eqref{eq_y_lin},
	\begin{align}\nonumber
		\mb{e}_{i}(k) &= F\mb{x}(k-1)+G\mb{w}(k-1)  -  \sum_{j\in i \cup \mathcal{N}_i} W_{ij}F\widehat{\mb{x}}_j(k-1) \nonumber \\
		&- K_{i} H_i^\top(H_i\mb{x}(k) + \bm \nu_i(k) - H_i\sum_{j\in i \cup \mathcal{N}_i} W_{ij}F\widehat{\mb{x}}_j(k-1) ) \nonumber \\ \nonumber
		&=
		F\mb{x}(k-1) -
		\sum_{j\in i \cup \mathcal{N}_i} W_{ij}F\widehat{\mb{x}}_j(k-1)
		\nonumber \\
		\nonumber
		&- K_{i} H_i^\top (H_iF\mb{x}(k-1) - H_i\sum_{j\in i \cup \mathcal{N}_i} W_{ij}F\widehat{\mb{x}}_j(k-1) )  \\
		&+ G\mb{w}(k-1)- K_{i} (H_i^\top H_iG\mb{w}(k-1)+ H_i^\top \bm \nu_i(k)).
		\nonumber
	\end{align}
	Collecting the noise terms in variable~$\bm \eta_i(k)$,
	\begin{align}
		\mb{e}_{i}(k) &=
		F\mb{x}(k-1) -
		\sum_{j\in i \cup \mathcal{N}_i} W_{ij}F\widehat{\mb{x}}_j(k-1)
		\nonumber \\
		&- K_{i} H_i^\top H_i(F\mb{x}(k-1) - \sum_{j\in i \cup \mathcal{N}_i} W_{ij}F\widehat{\mb{x}}_j(k-1) ) \nonumber \\&+\bm \eta_i(k).
		\nonumber
	\end{align} \normalsize
	Applying the row-stochasticity of~$W$~matrix,
	\begin{align}
		\mb{e}_{i}(k) &=
		\sum_{j\in i \cup \mathcal{N}_i} W_{ij}F(\mb{x}(k-1)-\widehat{\mb{x}}_j(k-1))
		\nonumber \\
		\nonumber
		&- K_{i} H_i^\top H_i\sum_{j\in i \cup \mathcal{N}_i} W_{ij}F(\mb{x}(k-1)-\widehat{\mb{x}}_j(k-1))
		\\ \nonumber &+\bm \eta_i(k).
	\end{align} \normalsize
	and
	Eq.~\eqref{eq_err1} follows. This completes the proof.	
\end{proof}
\textbf{Proof of Lemma~\ref{lem_obsrv}:}
\begin{proof}
	The proof follows from standard dynamic system stability analysis (e.g., Kalman stability theorem) \cite{bay}. Based on Kalman, the error dynamics in the following form (with $\mb{e}_c$ as the centralized estimation error, $A,H$ as the system and output matrices, and $\bm \eta$ as the noise term),
	\begin{align}
		\mb{e}_c(k) = \left(A - K H A\right)\mb{e}_c(k-1)+
		\bm \eta(k),
	\end{align}
	is stabilizable (i.e., there exists feedback gain matrix $K$ to make it stable) if the pair $(A,H)$ is observable. Similarly, for stabilizability of the (distributed) error dynamics \eqref{eq_err1} the equivalent pair $(W \otimes F,D_H)$ must be observable to ensure the existence of an observer/estimator gain $K$ to make the error dynamics stable. The algorithm to design such a \textit{block-diagonal} $K$ matrix to meet the distributed and localized property of the estimator is given in the existing literature, e.g., see \cite{rami:97}.
\end{proof}
\textbf{Proof of Lemma~\ref{lem_rank}:}
\begin{proof}
	All the diagonal entries of the transition matrix $F$, in \eqref{F_ncv} and \eqref{F_nca}, are non-zero. Since these non-zero diagonal entries share no rows and columns, they represent the set of (self) links in the associated digraph $\mc{G}_F$ forming a maximum matching which \textit{spans all nodes}. This property implies a full G-rank matrix $F$.  This can be checked numerically as the determinant of upper-triangular $F$ matrices in \eqref{F_ncv} and \eqref{F_nca} are equal to $1$ and both are full-rank (for all values of $T$).
\end{proof}
\textbf{Proof of Lemma~\ref{lem_parent}:}
\begin{proof}
	The proof is similar to that in \cite{pequito2015framework,jstsp14}. Recall that for structural observability two conditions hold \cite{pequito2015framework,jstsp14}: (i) full structural rank condition (which holds from Lemma~\ref{lem_rank}), and (ii) the accessibility condition, which implies that there exists a path from every state node to an output. Since a parent SCC/node has no outgoing link to other SCCs/nodes (from its definition), condition (ii) is only satisfied if there is a direct output from the parent node or one state node in the parent SCC. Since all nodes in the same parent SCC are strongly connected and for all the Child nodes/SCCs there is a directed path ending-up in a parent node/SCC, the observability condition (ii) is satisfied. This completes the proof.
\end{proof}
\textbf{Proof of Theorem~\ref{thm_sc}:}
\begin{proof}
	The proof follows from sufficient conditions for the Kronecker product observability of $W \otimes F, D_H$ and the concept of \textit{Kronecker network product}. The sufficient conditions on the observability of Kronecker network product are recalled here from \cite{TSIPN20}. Given the full G-rank system digraph $\mc{G}_F$ with sufficient measurements for observability (Lemmas~\ref{lem_rank} and~\ref{lem_parent}), the \textit{strong-connectivity} of $\mc{G}_W$ guarantees the structural observability of the composite network $\mc{G}_W \otimes \mc{G}_F$ (the Kronecker network product of $\mc{G}_W$ and $\mc{G}_F$). This is the minimal sufficient condition for network observability guarantee.
	Note that observability of the Kronecker network $\mc{G}_W \otimes \mc{G}_F$ is equivalent with $(W \otimes F, D_H)$-observability (in the \textit{structural} sense). This holds for any cyclic system digraph including both NCV and NCA target models and the theorem follows.
\end{proof}
\textbf{Proof of Theorem~\ref{thm_tau}:}
\begin{proof}
	Proof follows from \cite[Lemma~5]{LCSS_delay}.
	We need to show that, given Schur stability of   $\rho({\widehat{F}})<1$, its \textit{augmented version} $\underline{\widehat{F}} := W\otimes F^{\tau+1} - K \overline{D}_H (W\otimes F^{\tau+1})$ (the closed-loop error matrix associated with the delayed case) satisfies $\rho(\underline{\widehat{F}})<1$ for  $\tau<\overline{\tau}$. Recall that $\rho({\widehat{F}})<1$ implies that the sensors track the target via \eqref{eq_p}-\eqref{eq_m} with bounded steady-state error in the \textit{absence of delays}. Note that $\rho({F})^{\overline{\tau}+1}=\rho({F})=1$ and, from \cite[Lemma~5]{LCSS_delay}, for the closed-loop \textit{augmented} matrix  $\underline{\widehat{F}}$  we have,
	\begin{align} \label{eq_tau*2}
		\rho(\underline{\widehat{F}})\leq \rho(W\otimes F^{\overline{\tau}+1} - K \overline{D}_H (W\otimes F^{\overline{\tau}+1}) )^{\frac{1}{\overline{\tau}+1}},
	\end{align}
	which implies that $\rho(\underline{\widehat{F}})<1$ for some $\overline{\tau}$ since  $\rho(\widehat{F})<1$ for the designed block-diagonal $K$ matrix. Since the right-hand-side of Eq.~\eqref{eq_tau*2} is increasing on $\overline{\tau}$ (for $\rho(\widehat{F})<1$ and $\rho(F)=1$), if the Schur stability holds, for say a given $\overline{\tau}$, then it holds for any $\tau \leq \overline{\tau}$. In other words, Schur stability of $\widehat{F}$ also ensures the stability of $\underline{\widehat{F}}$ for all $\tau \leq \overline{\tau}$, and the proof follows.
\end{proof}

\section*{Acknowledgements} \label{sec_con}
This work is supported in part by the European Commission through the H2020 Project Finest Twins under grant agreement 856602 and H2022 project MINERVA under grant agreement 101044629.

\bibliographystyle{elsarticle-num}
\bibliography{bibliography}
\end{document}